\documentclass[aps,twocolumn,showpacs,superscriptaddress]{revtex4}
\usepackage{epsfig}
\usepackage{color}
\usepackage{bm}

\def\ketm#1{  \left\vert  #1   \right\rangle   }

\def\sprm#1#2{  \left\langle #1 \left\vert \right. #2 \right\rangle   }

\def\mem#1#2#3{  \left\langle #1 \left\vert  #2 \right\vert #3 \right\rangle   }

\begin{document}

%
%
%
%

\title{Inter--electronic interaction effects on the polarization of recombination photons}

%
%
%
%

%
%
%
%

\author{A.~Surzhykov\footnote{Corresponding author.
Email: surz@physi.uni-heidelberg.de}}

\affiliation{Physikalisches Institut, Universit\"{a}t Heidelberg,
Philosophenweg 12, D--69120 Heidelberg, Germany}
\affiliation{GSI Helmholtzzentrum f\"ur Schwerionenforschung GmbH, Planckstrasse 1,
D--64291 Darmstadt, Germany}

\author{A.~N.~Artemyev}

\affiliation{Physikalisches Institut, Universit\"{a}t Heidelberg,
Philosophenweg 12, D--69120 Heidelberg, Germany}
\affiliation{GSI Helmholtzzentrum f\"ur Schwerionenforschung GmbH, Planckstrasse 1,
D--64291 Darmstadt, Germany}

\author{V.~A.~Yerokhin}

\affiliation{Physikalisches Institut, Universit\"{a}t Heidelberg,
Philosophenweg 12, D--69120 Heidelberg, Germany}
\affiliation{GSI Helmholtzzentrum f\"ur Schwerionenforschung GmbH, Planckstrasse 1,
D--64291 Darmstadt, Germany}

\affiliation{St.~Petersburg State
Polytechnical University, Polytekhnicheskaya 29,
St.~Petersburg 195251, Russia}

\date{\today}

%
%
%
%
%
%
\begin{abstract}
A theoretical investigation of the radiative capture of an electron into a bound state of heavy, hydrogen--like ion is presented. Special attention is paid to the question of how the linear polarization of the emitted radiation is affected by the inter--electronic interaction effects. An analysis of these effects is performed within both, the screening--potential approximation and the perturbation theory that treats rigorously the electron correlations to the first order in the parameter 1/$Z$. By making use of these two approaches, detailed calculations are performed for relativistic collisions of hydrogen--like europium Eu$^{62+}$, bismuth Bi$^{82+}$ and uranium U$^{91+}$ ions with free electron and low--$Z$ atomic targets. Results of the calculations indicate that the two--electron effects may significantly influence the polarization properties of the recombination x--rays; the effect which can be observed by the present--day polarization detectors.
\end{abstract}


\maketitle

%
%
%
%

\section{Introduction}

In recent years, significant progress has been made in the development of segmented Ge and Si(Li) detectors \cite{Spi08,Web10,Tas06}. By exploiting the polarization sensitivity of the Compton effect, these detectors allow efficient linear--polarization studies of x--rays with energies ranging from tens to hundreds of keV. Both the degree and direction of such a polarization can be determined with a high accuracy, thus opening up new opportunities for a detailed investigation of the hard--photon emission from heavy atomic systems. At the GSI facility in Darmstadt, in particular, the solid--state Compton polarimeters have been successfully employed to explore the (polarization) properties of the photons accompanying the electron transfer from a target into bound states of highly--charged heavy ions \cite{Tas06,EiS07,WeB10}. Such a \textit{radiative recombination} (RR) of a free electron, known also as the \textit{radiative electron capture} (REC) of a loosely bound one, attracts much attention since its analysis reveals important knowledge about the electron–-photon interaction in the presence of extremely strong electromagnetic fields.

Until now, the RR (REC) polarization studies in the high--$Z$ domain have mainly dealt with the initially \textit{bare} ions. However, the future experiments, planed for the GSI and the international Facility for Antiprotons and Ion Research (FAIR), are likely to be focused on the electron capture by \textit{few--electron} species \cite{FAIR}. These measurements are required, for example, to explore atomic parity--violation (APV) phenomena in heavy atomic systems \cite{MaP09} or to diagnose heavy ion beams at storage rings. In the latter case, the rotation of the linear polarization of $K$--RR photons out of the reaction plane may provide unique information on the longitudinal polarization of (initially) hydrogen--like projectiles \cite{SuF05}. Such a polarization sensitivity has attracted a lot of recent interest since the operational access and control of polarized heavy ions is a stringent requirement for performing atomic tests of the standard model \cite{LaN01,Jun03}.

Any accurate polarization analysis of the x--ray emission accompanying the recombination of few--electron ions requires knowledge of the \textit{interelectronic--interaction} effects, whose importance was emphasized recently in Ref.~\cite{ReG08}. The influence of these effects on the RR properties has been discussed in a number of theoretical works \cite{PrR73,Sco95,PrL95,LaP98,TrN03,Wei05,FrS05}. In most previous studies, however, the $e$--$e$ effects have been treated in the screening--potential approximation based on the Dirac--Fock or Dirac--Slater calculations. Very recently, the accuracy of the screening model for the description of the RR of high--$Z$ ions has been questioned by us \cite{YeS10}. In particular, we argued that the electron correlation on the bound state together with the so--called off-resonant dielectronic recombination term (involving photon emission by a core electron) may significantly affect the total and angular--differential cross sections, yielding corrections comparable to those obtained in the screening--potential model. No attempt was done, however, to analyze how these corrections affect the polarization properties of RR photons.

In this contribution, therefore, we report a theoretical study of the electron--electron interaction effects on the linear polarization of the recombination photons. In our work, we shall focus on the experimentally relevant case of an electron capture by the (initially) hydrogen--like ions with energies of few hundreds of MeV/u. The theory required for the description of such a process is briefly outlined in Section \ref{sec_theory}. In particular, we provide the basic formulae for the linear polarization of the emitted light and discuss how it is parameterized in terms of the Stokes parameters. We show that any analysis of these parameters can be traced back to the (two--electron) matrix elements describing the free--bound electron transition under the simultaneous photon emission. The computation of these matrix elements, based on either the screening--potential approach or the perturbation theory, which---to the first order in the parameter 1/$Z$---accounts rigorously for the electron--electron interactions, is presented in Section \ref{sub_sec_matrix_element}. Together with the zero--order approximation, that neglects any $e$--$e$ correlations, these two ``many--electron'' theories are used in Section \ref{sec_results} to calculate the polarization parameters of the recombination photons. Here, we pay special attention to the electron capture into the ground state of the \textit{spin--polarized} medium-- and high--$Z$ ions whose polarization is known to give rise to a rotation of the RR linear polarization out of the reaction plane \cite{SuF05}. Based on the results of our calculations, we argue that $e$--$e$ interactions and, especially, the correlations beyond the screening model may influence the rotation (tilt) angle by about 10--20 \%; the effect which can be observed by the present--day polarization detectors. This effect is also briefly compared to the one arising from the binding of the target electrons, if radiative electron capture from \textit{atomic} targets is considered instead of radiative recombination of \textit{free} electrons. Summary of these results and their implications for the future x--ray polarization studies are given finally in Section \ref{sec_summary}.

%
%
%
%

\section{Theoretical background}
\label{sec_theory}

During the last decade, theoretical analysis of the radiative electron capture has been presented in detail in a number of publications \cite{EiS07,SuF05,ArS10,YeS10}. In what follows, therefore, we will restrict ourselves to a rather short account of the basic formulas, just enough to discuss the role of inter--electronic effects on the polarization of the recombination light. To start our brief review, we will define in the next subsection the Stokes parameters and will clarify their relation to the photons' density matrix.

\subsection{Polarization parameters and the density matrix}
\label{sub_sec_density_matrix}

>From an experimental viewpoint, the linear polarization of the emitted photons are most naturally described in terms of the Stokes parameters. These parameters are determined by the intensities of light $I_{\chi}$, polarized at different angles $\chi$ with respect to the \textit{reaction plane} that is spanned by the directions of the incident beam and emitted x--rays. While the parameter $P_1$ is obtained from the intensities of light, polarized in parallel and perpendicular to the reaction plane, the parameter $P_2$ follows from a similar intensity ratio, taken at $\chi = 45^\circ$ and $135^\circ$:
\begin{equation}
   \label{stokes_parameters}
   P_1 = \frac{I_{0^\circ} - I_{90^\circ}}{I_{0^\circ} + I_{90^\circ}} \, , \, \, \, \, \,
   P_2 = \frac{I_{45^\circ} - I_{135^\circ}}{I_{45^\circ} + I_{135^\circ}} \, .
\end{equation}
The Stokes parameters $P_1$ and $P_2$ can be easily ``visualized'' by means of the polarization ellipse, defined in the plane perpendicular to the photon momentum. The principal axis of such an ellipse is characterized by its tilt angle $\chi_0$ with respect to the reaction plane:
\begin{equation}
   \label{tilt_angle}
   \tan 2\chi_0 = \frac{P_2}{P_1} \, ,
\end{equation}
and by the relative length, $P_L = \sqrt{P_1^2 + P_2^2}$, which is attributed to the degree of linear polarization. When employing novel position--sensitive solid--state detectors, both, the degree $P_L$ and the tilt angle $\chi_0$ of recombination light, can be uniquely restored from the angular distributions of the Compton--scattered photons \cite{Tas06,Spi08,Web10}.

The use of the Stokes parameters (\ref{stokes_parameters}) is also very convenient for the \textit{theoretical} analysis of the RR polarization. These two parameters, together with the degree of circular polarization $P_3$,  are directly related to the density matrix of the emitted light:
\begin{eqnarray}
   \label{density_matrix}
   \left( \mem{{\bm k} \bm{u}_\lambda}{\hat{\rho}_\gamma}{{\bm k} \bm{u}_{\lambda'}} \right)_{\lambda,\lambda'=\pm1} =
   \;\frac{1}{2}\;\left( \begin{array}{cc}
   1 + P_3& P_1 - iP_2\\
   P_1 + iP_2&1 - P_3
\end{array} \right) .
\end{eqnarray}
Here, $\lambda = \pm 1$ is the helicity of the photon (i.e., the spin projection onto the direction of propagation) and ${\bm k}$ denotes its momentum. The density matrix on the left--hand side of Eq.~(\ref{density_matrix}) depends, of course, on the spin states of the particles involved in the collision and on the set--up under which the polarization properties of recombination photon are ``measured''. For the capture of unpolarized electrons into a (final) ionic level $\ketm{\alpha_f J_f}$, whose magnetic substates remain unobserved in a particular experiment, the elements of the density matrix read, for example, as:
\begin{eqnarray}
   \label{density_matrix_final}
   \mem{{\bm k} \bm{u}_\lambda}{\hat{\rho}_\gamma}{{\bm k} \bm{u}_{\lambda'}} &=& \frac{1}{2} \sum\limits_{M_f \, m_s} \, \sum\limits_{M_i M'_i} \, \mem{\alpha_i J_i M_i}{\hat{\rho}_i}{\alpha_i J_i M'_i} \nonumber \\
   & & \hspace*{-1cm} \times \mem{{\bm k} \bm{u}_\lambda, \alpha_f J_f M_f}{\hat{\mathcal O}}{{\bm p} m_s, \alpha_i J_i M_i} \nonumber \\
   & & \hspace*{-1cm} \times \mem{{\bm k} \bm{u}_{\lambda'}, \alpha_f J_f M_f}{\hat{\mathcal O}}{{\bm p} m_s, \alpha_i J_i M'_i}^* \, .
\end{eqnarray}
In this expression, ${\bm p}$ and $m_s$ are the asymptotic momentum and spin projection of the incident electron, the density operator $\hat{\rho}_i$ described the ion in its initial state $\ketm{\alpha_i J_i}$, and $\hat{\mathcal O}$ is the transition operator. Moreover, apart of the total angular momenta $J_{i,f}$ and their projections $M_{i,f}$, the $\alpha_{i,f}$ are used to denote all the additional quantum numbers as needed for a unique specification of the (many--electron) ionic states.

As seen from Eqs.~(\ref{density_matrix})--(\ref{density_matrix_final}), any further analysis of the polarization Stokes parameters requires evaluation of the amplitudes that describe the free--bound electron transition under a simultaneous emission of the recombination photon. To proceed with such an evaluation, we shall agree first about explicit form of the (many--electron) wavefunctions as well as of the operator $\hat{\mathcal O}$. This issues will be discussed in the next subsection.

\subsection{Transition matrix element}
\label{sub_sec_matrix_element}

No assumptions about the shell structure of the initial-- and the final--state ion have been made in the derivation of Eq.~(\ref{density_matrix}) which displays, therefore, the most general form of the photon spin--density matrix. In the present work, we shall employ such a matrix for the analysis of the polarization properties of the x--ray emission accompanying the electron capture into \textit{hydrogen--like} ions. The (initial) bound states of these ions are described by the well--known single--electron wavefunctions, $\ketm{\alpha_i J_i M_i} \equiv \ketm{n_i \kappa_i \mu_i}$. In contrast, some approximate methods are required to evaluate the two-electron states $\ketm{\alpha_f J_f}$, following the electron recombination. Here, the helium--like wavefunctions are constructed within the single--determinant approach:
\begin{eqnarray}
   \label{he_like_function}
   \Psi_{\alpha_f J_f M_f}({\bm r}_1, {\bm r}_2) &=& N \, \sum\limits_{\mu'_i \mu_b} \, \sprm{j_i \mu'_i \, j_b \mu_b}{J_f M_f} \nonumber \\
   &\times&
   \left| \matrix{ \phi_{n_i \kappa_i \mu'_i}({\bf r}_1) & \phi_{n_b \kappa_b \mu_b}({\bf r}_1)  \cr
                   \phi_{n_i \kappa_i \mu'_i}({\bf r}_2) & \phi_{n_b \kappa_b \mu_b}({\bf r}_2)  } \right| \, ,
\end{eqnarray}
which accounts for the Pauli principle and the proper coupling of the angular momenta. As usual, the normalization constant is chosen to be $N = 1/2$ for equivalent electrons and $N = 1/\sqrt{2}$ otherwise. Moreover, as seen from Eq.~(\ref{he_like_function}), we assume that the initially bound electron in the state $\ketm{n_i \kappa_i}$ does not undergo a transition to a higher (lower) level even though its spin projection can be changed in course of the capture process.

\begin{figure}
\centerline{\includegraphics[width=0.5\textwidth]{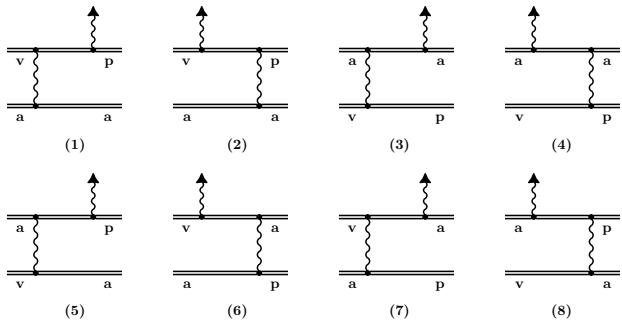}}
\caption{The one--photon exchange correction to the transition amplitude of the radiative recombination of an electron with a hydrogen--like ion.
The double line indicates an electron propagating in the field of a nucleus. The wavy line with an arrow denotes the emitted photon. The incoming
electron is denoted as $p$, $a$ is the initially bound (spectator) electron, and $v$ is the captured electron.
\label{fig:1phot}}
\end{figure}

Any practical use of the final--state wavefunction (\ref{he_like_function}) requires the knowledge of the orbitals $\phi_{n \kappa \mu}$. In the simplest (zero--order) approximation, these orbitals can be taken as the well--known solutions of the single--electron Dirac Hamiltonian. Such an independent particle model allows significant simplification of the amplitudes in Eq.~(\ref{density_matrix}) if the transition operator $\hat{\mathcal O}$ is supposed to act locally on either of two electrons, leaving the other one unchanged:
\begin{equation}
   \label{operator_model}
   \hat{\mathcal O} = \hat{\mathcal R}_1 \otimes \hat{I}_2 + \hat{\mathcal R}_2 \otimes \hat{I}_1 \, .
\end{equation}
That is, by inserting Eqs.~(\ref{he_like_function}) and (\ref{operator_model}) into the transition matrix element from the second line of Eq.~(\ref{density_matrix}) we easily derive:
\begin{eqnarray}
   \label{matrix_element_ipm}
   \mem{{\bm k} \bm{u}_\lambda, \alpha_f J_f M_f}{\hat{\mathcal O}}{{\bm p} m_s, \alpha_i J_i M_i} & & \nonumber \\
   && \hspace*{-5.5cm} \equiv \mem{{\bm k} \bm{u}_\lambda, (n_i \kappa_i, n_b \kappa_b) J_f M_f}{\hat{\mathcal O}}{{\bm p} m_s, n_i \kappa_i \mu_i}
   \nonumber \\[0.1cm]
   && \hspace*{-5.5cm} = N \sum\limits_{\mu'_i \mu_b} \sprm{j_i \mu'_i \, j_b \mu_b}{J_f M_f} \nonumber\\
   && \hspace*{-5.5cm} \times \left( \delta_{\mu_i \mu'_i} \, \tau^{(0)}_{m_s \mu_b}
   - \delta_{\mu_i \mu_b} \delta_{n_i n_b} \delta_{\kappa_i \kappa_b} \, \tau^{(0)}_{m_s \mu'_i}
   \right) \, , \nonumber \\
\end{eqnarray}
where, by following Ref.~\cite{YeS10}, we introduced a notation:
\begin{equation}
   \label{matrix_element_one_electron}
   \tau^{(0)}_{m_s \mu_b} = \mem{n_b \kappa_b \mu_b}{{\bf \alpha} {\bf u}^*_{\lambda} {\rm e}^{-i {\bf kr}} }{{\bm p} m_s} \, ,
\end{equation}
for the one--particle RR amplitude which have been applied very frequently in the past for studying the (photo--) ionization and electron capture processes \cite{EiS07,PrR73}. In this amplitude, the standard operator $\hat{\mathcal R} = {\bf \alpha} {\bf u}_{\lambda} {\rm e}^{i {\bf kr}}$ describes the interaction of an electron with the radiation field.

The two--electron matrix elements (\ref{matrix_element_ipm})--(\ref{matrix_element_one_electron}), being derived in the zero--order model, obviously can not be used for the proper analysis of the inter--electronic effects on the RR polarization. In the following, therefore, we shall discuss other approaches that may account for the (major part of) $e-e$ interactions. Within the high--$Z$ domain, the \textit{screened--potential approximation} provide an effective tool to explore the radiative recombination by initially hydrogen--like ions. The great advantage of such an approximation is that it still allows the decomposition of the helium--like transition amplitudes in terms of their one--electron analogs (cf. Eq.~(\ref{matrix_element_ipm})). In contrast to the independent particle model, however, the bound as well as the continuum single--electron orbitals are obtained in this case by solving the Dirac equation with the screening potential of the core electron. In fact, this treatment is equivalent to the frozen--core Dirac--Fock method (as the core in our case contains only one electron).

In order to go \textit{beyond} the screening--potential approximation, one has to use the rigorous quantum electrodynamic (QED) formalism. Within such a formalism, the electron--electron interaction is described by the exchange of (an arbitrary number of) the virtual photon(s). For heavy ions, it is sufficient to restrict the analysis to the one--photon exchange only, as the exchange by two and more photons is suppressed by an additional factor of 1/$Z$. The gauge invariant set of Feynman diagrams representing the one--photon exchange corrections to the transition amplitude is depicted in Fig.~\ref{fig:1phot}. Denoting these corrections as $\tau^{(1,i)}$ ($i = 1\ldots 8$), we write the total amplitude as
\begin{eqnarray}
   \label{matrix_element_QED}
   \mem{{\bm k} \bm{u}_\lambda, \alpha_f J_f M_f}{\hat{\mathcal O}}{{\bm p} m_s, \alpha_i J_i M_i} & & \nonumber \\
   && \hspace*{-5.5cm} =  N \sum\limits_{\mu'_i \mu_b} \sprm{j_i \mu'_i \, j_b \mu_b}{J_f M_f}
   \nonumber \\
   && \hspace*{-5.5cm} \times \left( \delta_{\mu_i \mu'_i} \, \tau^{(0)}_{m_s \mu_b}
   - \delta_{\mu_i \mu_b} \, \delta_{n_i n_b} \delta_{\kappa_i \kappa_b}  \, \tau^{(0)}_{m_s \mu'_i}
   \right) \nonumber \\
   && \hspace*{-5.5cm} + N \sum\limits_{\mu'_i \mu_b} \sprm{j_i \mu'_i \, j_b \mu_b}{J_f M_f} \, \sum\limits_{i = 1}^8 \tau^{(1,i)}_{m_s \mu_i \mu_b \mu'_i} \, ,
\end{eqnarray}
where the second and third lines represent the unperturbed (zero--order) matrix element. We note that the first--order corrections, given in the last line of Eq.~(\ref{matrix_element_QED}), contain the summation over the complete Dirac spectrum and correspond to the effects that are omitted in the standard many--body techniques (in particular, the emission of radiation from the core electron and the {non-resonant} dielectronic recombination, see Ref.~\cite{YeS10} for details).

As seen from Eqs.~(\ref{density_matrix}), (\ref{density_matrix_final}) and (\ref{matrix_element_QED}), the perturbative analysis of inter--electronic
interaction effects on the RR linear polarization can be traced back to the evaluation of the zero-- as well as first--order transition amplitudes. The computation of $\tau^{(0)}_{m_s \mu_b}$, given by Eq.~(\ref{matrix_element_one_electron}), follows the standard procedure \cite{YeS10,SuF02} in which one just needs to care about the change of the transition energy due the screening of the nucleus by the spectator electron. In contrast, the treatment of the first--order amplitudes is a rather demanding task because it requires the evaluation of the electron propagators. In the present work, the numerical evaluation of $\tau^{(1,i)}$ is performed within the Dirac Coulomb Green's function approach. We shall not discuss this computation further here but rather refer for all further details to the literature \cite{YeS00,MIST,YeS10}.

\begin{figure}[t]
\begin{center}
\epsfig{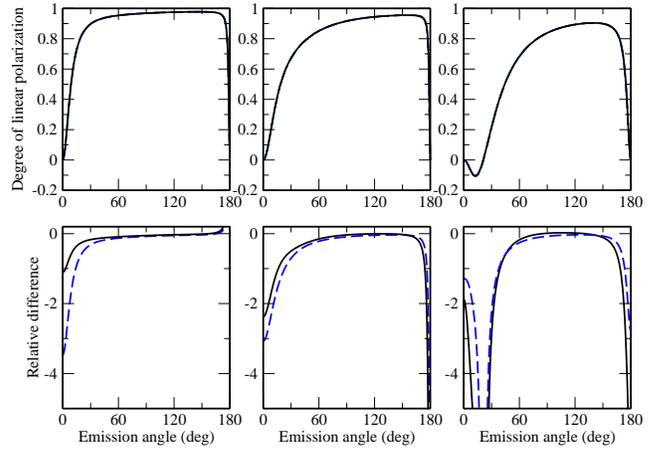} \caption{\label{fig1}
(Color online) In the top panel we display the degree of the linear polarization of the photons which are emitted in the radiative recombination of an electron into the ground state of (initially) hydrogen--like europium Eu$^{62+}$ ion with projectile energies of 100 (left column), 300 (middle column) and 600 MeV/u (right column). The difference between the predictions of the zero--order approximation, as well as the screening--potential-- and the perturbative approaches lays within the width of the lines used in the figure. In order to visualize the inter--electronic interaction effects, the \textit{relative} difference (\ref{difference}) between (i) the zero--order and screening--potential calculations (dashed line) and (ii) the zero--order and perturbative calculations (solid line) is presented in the bottom panel. Results are given in the laboratory frame.}
\end{center}
\end{figure}
%
%
%
%

%
%
%
%

\section{Results and discussion}
\label{sec_results}

\subsection{Radiative recombination of unpolarized ions}

\begin{figure}[t]
\begin{center}
\epsfig{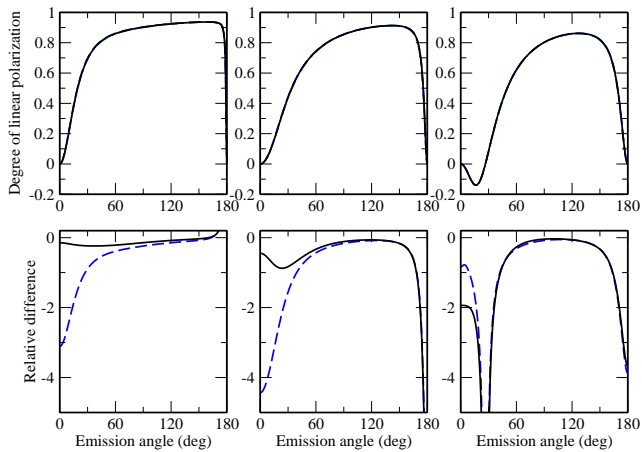} \caption{\label{fig2}
(Color online) In the top panel we display the degree of the linear polarization of the photons which are emitted in the radiative recombination of an electron into the ground state of (initially) hydrogen--like uranium U$^{91+}$ ion with projectile energies of 100 (left column), 300 (middle column) and 600 MeV/u (right column). The difference between the predictions of the zero--order approximation, as well as the screening--potential-- and the perturbative approaches lays within the width of the lines used in the figure. In order to visualize the inter--electronic interaction effects, the \textit{relative} difference (\ref{difference}) between (i) the zero--order and screening--potential calculations (dashed line) and (ii) the zero--order and perturbative calculations (solid line) is presented in the bottom panel. Results are given in the laboratory frame.}
\end{center}
\end{figure}

With the formalism developed above, we are ready now to analyze the electron--electron interaction effects on the linear polarization of the photons emitted in the radiative recombination of initially hydrogen--like ions. As seen from Eqs.~(\ref{density_matrix})--(\ref{density_matrix_final}), the polarization is affected also by the initial spin--state of the ion as described by the operator $\hat{\rho}_i$. Since most of today's experiments deal with the beams of \textit{unpolarized} heavy ions, we define first the matrix of this operator as:
\begin{equation}
   \label{initial_state_unpolarized}
   \mem{\alpha_i J_i M_i}{\hat{\rho}_i}{\alpha_i J_i M'_i} \, = \, \delta_{M_i M'_i} \, \frac{1}{2J_i + 1} = \, \delta_{M_i M'_i} \frac{1}{2} \, ,
\end{equation}
where we assume, moreover, the spectator electron to be in the $1s_{1/2}$ ground state. For such a ``preparation'' of the initial state, the second Stokes parameter $P_2$ vanishes identically and the photon's linear polarization is described by a single parameter $P_1$ (cf. Refs.~\cite{SuF05,SuF03}). In Figs.~\ref{fig1}--\ref{fig2}, we display this parameter as a function of the emission angle $\theta$ for the electron recombination into the $1s_{1/2}^2$ state of (finally) helium--like europium and uranium ions with energies $T_p$ = 100, 300 and 600 MeV/u. The theoretical predictions obtained within the relativistic perturbative theory (\ref{matrix_element_QED}) are compared here with the screening--potential calculations and the results of the zero--order approach (\ref{matrix_element_ipm}) that treats the electrons as independent. In order to emphasize the role of the inter--electronic correlations, we display in the bottom panel of Figs.~\ref{fig1}--\ref{fig2} also the relative difference:
\begin{equation}
   \label{difference}
   \Delta = \frac{P_1^{(0)} -  P_1}{\left| P_1^{(0)} \right|} \, \times \, 100 \% \, ,
\end{equation}
between the Stokes parameters as evaluated within the independent particle model, $P_1^{(0)}$, and within both (perturbative and screening--potential) ``many--electron'' approaches. As seen from the figures, the most pronounced electron--electron interaction effects can be observed for the forward and backward photon emission. We shall note however, that RR linear polarization identically vanishes for $\theta$ = 0$^\circ$ and 180$^\circ$ since the axial symmetry of the system ``ion + photon'' is preserved at these angles and, hence, no reaction plane can be uniquely defined. Moreover, for the large angles, $\theta > 160^\circ$, the Stokes parameter can hardly be measured due to the low photon intensity (see, e.g. \cite{EiS07,SuF05}). Therefore, the only angular range where the inter--electronic interactions may become ``visible'' in the present--day experimental studies is $5^\circ \lesssim \theta \lesssim 40^\circ$. In this region and for relatively low collision energies, $T_p \lesssim$ 400 MeV/u, the $e$--$e$ interactions result in a 1--2 \% enhancement of the polarization's degree $P_1$ which can be attributed to the partial screening of the nucleus by the spectator electron and to the corresponding reduction of the transition energy. For the same reason, the zero--order approach overestimates the \textit{crossover} effect that is observed at small emission angles, $\theta \lesssim$ 30$^\circ$, and projectile energies $T_p \gtrsim$ 500 MeV/u. This effect is reflected by the negative values of the Stokes parameter $P_1$ and is known to be more pronounced for the higher transition energies (see, e.g. Ref~\cite{EiS07,EiI02}).

Although \textit{qualitatively} predicting the enhancement of the linear polarization of $K$--RR photons, the screening--potential approximation yields results which are rather different from the perturbative calculations. As seen from the bottom panel of Figs.~\ref{fig1}--\ref{fig2}, this approximation significantly---up to factor of \textit{four}---overestimates the inter--electronic effects for the energies $T_p \lesssim$~300~MeV/u. In contrast, when the projectile energy rises to 600~MeV/u, the screening--potential model suggests that the polarization of x--rays, emitted in the forward directions, is weaker affected by the $e$--$e$ repulsion than predicted by the rigorous perturbation theory.

\subsection{Radiative recombination of polarized ions}

Until now we have discussed the polarization of RR x--rays that emerge in energetic collisions between \textit{unpolarized} ions and electrons. This scenario corresponds to the typical set--up in the present storage--ring studies \cite{Tas06}. However, in the future experiments that are planned to be performed at the Facility for Antiproton and Ion Research (FAIR) in Darmstadt, special attention shall be given to interactions of spin--polarized projectile ions and/or target atoms. The ``spin--dependent'' studies are expected to reveal unique information on the relativistic, many--body and QED effects on the structure and dynamics of heavy atomic systems. Furthermore, application of \textit{longitudinally} polarized heavy ion beams opens up a very promising way to explore atomic parity non--conservation (PNC) phenomena as well as to search for electric dipole moments of heavy nuclei as proposed as a test of the Standard Model. In order to efficiently produce such---polarized---beams an optical pumping of the hyperfine levels $\ketm{F M_F}$ of hydrogen--like species has been proposed by Prozorov and co--workers \cite{PrL03}. Since this hyperfine state results from the coupling of the electron and the nucleus, it requires that the nuclear spin be polarized as well. The key parameter in such a scenario is the so--called degree of ion polarization:
\begin{equation}
   \label{degree_of_ion_polarization}
   \lambda_F =  \frac{1}{F} \, \sum_{M_F} n_{M_F} M_F  \, ,
\end{equation}
where $n_{M_F}$ is the relative population of the hyperfine sublevel $\ketm{F M_F}$. We have argued recently that detailed information about this polarization can be obtained from the analysis of the Stokes parameters of RR radiation if unpolarized electrons are captured into the ground state of finally helium--like ions \cite{SuF05}. In particular, we found that the parameter $P_1$ does not depend on ion polarization, while the second Stokes parameter $P_2$ appears to be directly proportional to it, $P_2(\theta) \propto \lambda_F$. As seen from Eq.~(\ref{tilt_angle}) this implies rotation of the polarization ellipse out of the reaction plane on the angle \cite{SuF05}:
\begin{equation}
   \label{tilt_angle_polarization}
   \tan 2\chi_0 = \lambda_F \, \frac{I-1/2}{I+1/2} \, \mathcal{R}(T_p, Z) \, ,
\end{equation}
where $I$ is the nuclear spin and $\mathcal{R}$ is some function which depends only on collisional parameters such as the projectile's velocity and charge.

\begin{figure}[t]
\begin{center}
\epsfig{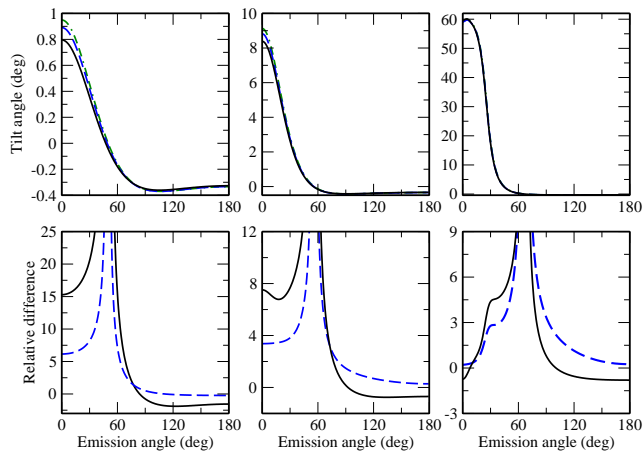} \caption{\label{fig3}
(Color online) In the top panel we display the tilt angle (\ref{tilt_angle}) of the linear polarization of the photons which are emitted in the radiative recombination of an electron into the ground state of (initially) polarized hydrogen--like europium Eu$^{62+}$ ion with projectile energies of 100 (left column), 300 (middle column) and 600 MeV/u (right column). Calculations, performed within the zero--order approximation (dash--and--dotted line) as well as in the screening--potential-- (dashed line) and perturbative (solid line) approaches, yield results that---except for low--energy collisions and small emission angles---can not be distinguished visually on the figure. In order to visualize the inter--electronic interaction effects, the \textit{relative} difference between (i) the zero--order and screening--potential calculations (dashed line) and (ii) the zero--order and perturbative calculations (solid line) is presented in the bottom panel. Results are given in the laboratory frame.}
\end{center}
\end{figure}

In Ref.~\cite{SuF05} we have employed the independent particle model in order to calculate the tilt angle (\ref{tilt_angle_polarization}) for the radiative recombination of (spin--polarized) hydrogen--like ions. No attempt was previously made, however, to analyze the role of the electron--electron interaction effects on such a rotation. In this contribution, we shall use both, the screening--potential and relativistic perturbative approximations from Sec.~\ref{sub_sec_matrix_element} to address the question of how the angle $\chi_0$ is affected by the $e$--$e$ correlations. In this line, we again start from Eqs.~(\ref{density_matrix})--(\ref{density_matrix_final}) where the density matrix of \textit{longitudinally--polarized} hydrogen--like ion is given now by \cite{SuF05,SuF03}:
\begin{eqnarray}
   \label{initial_state_polarized}
   \mem{\alpha_i J_i M_i}{\hat{\rho}_i}{\alpha_i J_i M'_i} \, &=& \, \delta_{M_i M'_i} \, \frac{1}{2} \nonumber \\
   & & \hspace*{-2cm} \times \left(1 + \lambda_F \, \frac{I-1/2}{I+1/2} \, (-1)^{1/2 - M_i} \right) \, .
\end{eqnarray}
In Figs.~\ref{fig3}--\ref{fig4} we display the tilt angle $\chi_0$ evaluated for such an initial--state density matrix. Calculations have been performed for the electron recombination into the ground state of completely polarized, $\lambda_F$ = 1, europium ($I$ = 5/2) and bismuth ($I$ = 9/2) ions with energies $T_p$ = 100, 300 and 600 MeV/u. Moreover, in the bottom panel of the figures we show the relative difference (defined similarly to Eq.~(\ref{difference})) between the predictions of the independent particle model and the screening--potential as well as perturbative approaches.

As seen from Figs.~\ref{fig3}--\ref{fig4}, the inter--electronic interactions lead to the reduction of the tilt angle $\chi_0$. The most pronounced effects are observed for for the forward emission angles, 5$^\circ \lesssim \theta \lesssim$ 30$^\circ$, where the difference between the predictions of the zero--order and ``two--electron'' theories is about 10--20~\% for the collision energy $T_p$ = 100 MeV/u and 2--5 \% for $T_p$ = 600 MeV/u. We note that the relative difference can reach even higher values if the tilt angle $\chi_0$ becomes very small, $\left| \chi_0 \right| \lesssim$ 0.5$^{\circ}$. This region, however, is not of interest for the present--day experimental studies owing to the limited (angular) resolution of the solid--state polarization detectors.

A remarkable reduction of the rotation angle $\chi_0$ of the $K$--RR linear polarization shall be partially attributed to the screening of the ionic nucleus by the spectator electron. Besides the modification of the initial-- and final--state wavefunctions (see Refs.~\cite{PrL95,LaP98}) this screening results in the decrease of the transition energy and, hence, of the second Stokes parameter $P_2$ (and, consequently, $\chi_0$). A further lowering of the tilt angle $\chi_0$ may be caused by the electron--electron correlations beyond the screening--potential approximation. In the experimentally relevant region of small emission angles the effect of these correlations is rather remarkable for relatively low collision energies but but becomes of minor importance for $T_p = $~600~MeV/u.

\subsection{Radiative electron capture by polarized ions}

\begin{figure}[t]
\begin{center}
\epsfig{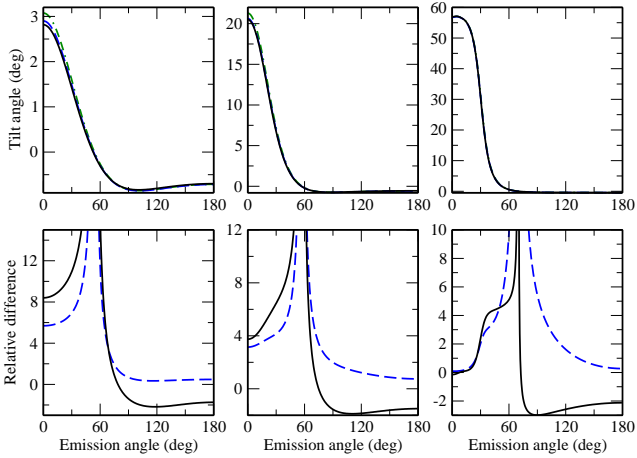} \caption{\label{fig4}
(Color online) In the top panel we display the tilt angle (\ref{tilt_angle}) of the linear polarization of the photons which are emitted in the radiative recombination of an electron into the ground state of (initially) polarized hydrogen--like bismuth Bi$^{82+}$ ion with projectile energies of 100 (left column), 300 (middle column) and 600 MeV/u (right column). Calculations, performed within the zero--order approximation (dash--and--dotted line) as well as in the screening--potential-- (dashed line) and perturbative (solid line) approaches, yield results that---except for low--energy collisions and small emission angles---can not be distinguished visually on the figure. In order to visualize the inter--electronic interaction effects, the \textit{relative} difference between (i) the zero--order and screening--potential calculations (dashed line) and (ii) the zero--order and perturbative calculations (solid line) is presented in the bottom panel. Results are given in the laboratory frame.}
\end{center}
\end{figure}

As seen from the results presented in Figs.~\ref{fig3}--\ref{fig4}, application of the RR for the diagnostics of spin--polarized heavy--ion beams requires, in general, an accurate analysis of the $e$--$e$ interactions. Up to now, we have discussed the recombination of a \textit{free} electron and, hence, examined these interactions as arising between the (continuum and bound) electrons exposed to the Coulomb field of the projectile nucleus. If, however, an incident electron is not free but bound to a low--$Z$ atom, which is typical for the present--day experiments, the target effects should also be taken into account. Recently, we have employed the impulse approximation based on the Roothaan--Hartree--Fock theory in order to study the influence of such effects on the polarization properties of the REC photons emitted in energetic collisions of bare projectiles with unpolarized targets \cite{ArS10}. Here, we extend this approach to explore the radiative electron capture by spin--polarized hydrogen--like ions. In the top panel of Fig.~\ref{fig5}, for example, we display the tilt angle (\ref{tilt_angle_polarization}) of the polarization ellipse of the $K$--REC photons as evaluated for collisions of (completely polarized) Bi$^{82+}$ projectiles with neutral helium, argon and krypton atoms. To illustrate the role of the target effects, the relative difference between the predictions obtained for the capture of a free electron (radiative recombination) and the REC calculations, $\Delta = (\chi_0^{RR} - \chi_0^{REC})/\left| \chi_0^{RR} \right|$, is presented also in the bottom panel. As seen from this figure, the tilt angle can be significantly reduced by the effects which arise from the binding of the target electrons. As could be expected, the greater difference between the RR and REC calculations is observed for heavier targets. For the electron capture from the neutral krypton atom, for example, the target effects may lead to a 10--20 \% reduction of the polarization rotation angle $\chi_0$. In contrast, in relativistic collisions with lightest atoms, the target effects are rather small and do not exceed those arising from the interaction with a projectile electron (cf. Fig.~\ref{fig4}).

%
%
%
%

%
%
%
%

\section{Summary}
\label{sec_summary}

\begin{figure}[t]
\begin{center}
\epsfig{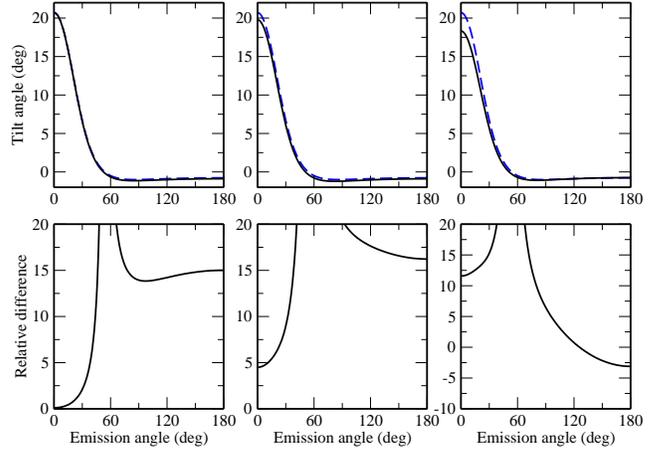} \caption{\label{fig5}
(Color online) In the top panel we display the tilt angle (\ref{tilt_angle}) of the linear polarization of the photons which are emitted in the radiative electron capture into the ground state of (initially) polarized hydrogen--like bismuth Bi$^{82+}$ ion with projectile energy 300 MeV/u. Calculations have been performed for the collisions with neutral helium (left column), argon (middle column) and krypton (right column) atoms.
Apart from the REC results (solid line), predictions are also presented for the linear polarization of radiative recombination photons (dashed line). The relative difference between the RR and REC calculations is given in the bottom panel. Results are presented in the laboratory frame.}
\end{center}
\end{figure}

In conclusion, we have investigated the radiative capture of an electron into a bound state of (initially) hydrogen--like heavy ions. In our theoretical analysis, we focus especially on the question of how inter--electronic interactions affect the polarization properties of the emitted x--rays. To account for the $e$--$e$ correlations, two models have been implemented: (i) the screening--potential approximation which is equivalent to the frozen--core Dirac--Fock method, and (ii) the perturbative theory. The latter allows to treat the inter--electronic effects \textit{rigorously} to the first order in the parameter 1/$Z$. For the capture into the ground state of medium-- and high--$Z$ ions, both theoretical approaches predict an enhancement of the (degree of) photon linear polarization. Such an enhancement is most sizable for the forward emission angles where the electron--interaction effects amount to 1--2 \%. When analyzing these effects, we stressed the role of the $e$--$e$ correlations \textit{beyond} the screening--potential model. These correlations contribute significantly to the overall ``many--electron'' effect and, hence, have to be taken into account for the accurate RR polarization calculations.

Besides the RR of unpolarized ions, we have also discussed the electron capture by longitudinally--polarized hydrogen--like species. In this case, the rotation angle $\chi_0$ of the $K$--RR polarization ellipse may serve as a \textit{measure} of the beam spin--polarization. We found that such a tilt angle can be significantly reduced by the electron--electron interactions. Apart from the screening corrections, this reduction should be partially attributed to the electron correlation on the bound electron state as well as to the off--resonance dielectronic recombination term. The latter two are especially important for slow collisions but become of less significance as the projectile energy increases.

Based on the results presented in Figs.~\ref{fig3}--\ref{fig4}, we argue that the electron--interaction corrections to the tilt angle $\chi_0$ are large enough to be measured by the present solid--state polarization detectors. Accurate treatment of these corrections are needed, therefore, for the precise diagnostics of the spin--polarized ion beams in storage rings as proposed in Ref.~\cite{SuF05}. However, such a diagnostics may be affected also by the $e$--$e$ correlations as occur in the target atom if an electron is initially bound to it. To estimate the size of the target effects we used the impulse approximation based on the Roothaan--Hartree--Fock theory \cite{ArS10} and calculated the tilt angle of the linear polarization of $K$--REC photons emitted in collisions of polarized hydrogen--like bismuth Bi$^{82+}$ with neutral target atoms. Results of our calculations, depicted in Fig.~\ref{fig5}, indicate that while for medium--$Z$ atoms the target effects are rather significant, they can be neglected in a low--$Z$ domain and for the forward emission angles favoring, thus, accurate measurements of the polarization of heavy ion beams.

%
%
%
%

\section*{Acknowledgments}
The work reported in this paper was supported by the Helmholtz Gemeinschaft (Nachwuchsgruppe VH--NG--421).

%
%
%
%


\begin{thebibliography}{99}
\bibitem{MIST} In the course of the present study, we found that our previous calculation \cite{YeS10} contained a mistake. Eq.~(2) of that work is
               valid for the capture into the excited states but not for the ground state. In the latter case, the zeroth-order amplitude should be antisymmetrized, see Eq.~(\ref{matrix_element_ipm}) of the present paper. As a result, the corrections $\sigma_{\rm corr}^{(1)}$ and $\sigma_{\rm DR}^{(1)}$ for the capture into the $(1s)^2$ state in Table II of Ref.~\cite{YeS10} should be multiplied by a factor of 2.
\end{thebibliography}
\end{document}